\def\beg{\begin{equation}}
\def\eeq{\end{equation}}
\documentstyle[12pt]{article}
\begin{document}
\begin{center}
{\Large{\bf Cyclotron Resonance of Composite Fermions:Quantum Hall Effect.}}
\vskip0.5cm
{\bf Keshav N. Shrivastava}\\
{\it School of Physics, University of Hyderabad,\\
Hyderabad  500 046, India.}
\end{center}
\vskip0.5cm
We have examined the claim made by Kukushkin et al that they have
 observed the cyclotron resonance of composite fermions. We find 
that the claim made is false and there is no justification for making
 false reports. The microwave absorption in cyclotron levels is obtained
 and it is claimed that CF has been seen. Since ``even flux quanta 
attachment to one electron" has not been seen, we find that Kukushkin's 
claim to have seen the CF is false. Since the attachment of the flux
 quanta to electrons can be an important discovery, Kukushkin et al 
have made the claim of seeing the CF without actually identifying them. 
The data can be interpretted without attaching flux quanta to the electrons.
\vskip5.0cm
keshav@mailaps.org   Fax. +91-40-3010145
\newpage

\noindent{\bf 1.~Introduction.}

First of all, we should define CF. By CF we mean composite 
fermions (CF) which are magnetic flux quanta attached to the electron.
 Even number of flux quanta, such as 2, are attached to one 
electron to make a CF. The effective field at the CF is given by,
\beg
B^* = B - 2n_o\phi_o\,\,\,,
\eeq
where $B$ is the exernal magnetic field, $n_s$ is the electron 
concentration per unit area and $\phi_o$ is the unit flux $\phi_o=hc/e$,
It is found by Jain [1-6] that attaching two flux quanta to one
 electron produces an effective charge of 1/3. Thus we have three
 properties to identify a CF. (1) Attachment of two flux quanta 
to an electron, (2) effective field  is B$^*$ and (3) the quasi 
particle charge is 1/3 for  the attachment of two flux to one electron.

We will now examine whether these properties are found by 
Kukushkin et al[7] or they have made a mistaken report. From 
this study we find that the identification of CF in the
 experimental data by Kukushkin et al is incorrect.

\noindent{\bf 2. The experimental measurement.}

Recently, Kukushkin et al[7] have measured the microwave 
absorption as a function of magnetic field. They find that
 absorption lines at the cyclotron frequency are symmetrically 
situated from the centre at $\nu $= 1/2 which occurs at a  field
 of about 9 $T$. The frequency as a function of field is a linear
 function. The effective field is identified by the expression (1).
 Substituting B$^*$ = 0 at $\nu $= 1/2, we find,
\beg
B = 2 n_s \phi_o
\eeq

We take $n_s$ = 1.09 $\times 10^{11}$ and $\phi_o$ = 4.13$\times 10 ^{-7}$
 Gauss cm$^2$ so that the field $B$ is found to be about 9 $T$. 
Alternatively, we take the field, at which $\nu $= 1/2 occurs and
 the value of $\phi_o$ to calculate the electron concentration, $n_s$.
 The field B as a function of $n_s$ is
approximately linear. However, upon examination of a few fundamental 
properties, it becomes clear that not everything is fine and there is
 something not quite correct.\\

\noindent{\bf I}. Near the B$^*$=0, the plot of microwave frequency
 as a function of the effective field B$^*$ is linear. Since $B=9T$, 
the plot of microwave frequency versus field, B$^*$, extends from 
negative to positive values. For $B< 2n_s \phi_o$, the value of 
$B^*$ is negative and for B$^* > 2n_s \phi_o$, it is positive. 
The graph appears symmetric near B$^* =0$. When we look at the 
values of the large plateaus, the right hand side which is the positive
 $B^*$ side, has a plateau at $\nu = 3/7$ whereas the left-hand 
side has 3/5. Basically it means that at the plateaus, the left-hand
 side is not equivalent to the right-hand side. If there is a time 
reversal invariance in the system, the right-hand side should be the
 same as the left hand side, like mirror images with mirror at $B^* =0$. On the
 other hand 3/5 on left and 3/7 on the right do not make the 
picture symmetric and the time reversal invariance is not obeyed 
by the experimental data. Therefore,  the data has not been identified
 correctly. Hence, eq.(1) which introduces two fields, one B and 
the other $B^*$ is not correct.When we pass a current in a coil,
 a field is produced. Reversing the direction of the current 
reverses the field. The expression (1) is obviously not having 
this  simple property and hence we can  say that the field 
discription(1) is not correct.\\
\noindent{\bf II}. We have shown[8] elsewhere that the usual
 Biot and Savarts law is not obeyed by the expression (1).
 According to which
 the current fully determines the field so there is no question of
 attaching the flux quanta to electrons. The flux quantization is
 determined by an expression of the type,
\beg
B. area = \phi_o
\eeq
but the expression (1) is not in agreement with the flux 
quantization. Therefore the idea of flux attachment to the electron
 is not correct. In any case, the data is asymmetric near $\nu=1/2$
 whereas the formula (2) is symmetric. Therefore, Kukushkin et al 
 have not varified the expression (1).\\
\noindent{\bf III}. According to the original papers [1-2] on the
 composite fermions the attachment of two flux quanta to one
 electron produces the fractional charge of $1/3$ but the experiment
 of Kukushkin et al has used the value of 1/2. Therefore,the 
experimental data does not agree with the CF value. The application
 of the two flux attachent to $\nu=1/2$ is not justified and hence
 the observed resonance is not the result of cyctotron resonance
 of CFs. The application of the two flux 
attachment to $\nu=1/2$ is not justified and hence the observed
 resonance is not the result of the cyclotron resonance of CFs. 
Different bands belong to different energies so that one 
obtains different effective mass for different frequencies,
 $\omega=eB/m^*$. Therefore, it is perfectly justified to use 
different masses in different bands but the effective charge
 identification by Kukushkin et al is mistaken. They have 
mistaken 1/2 for 1/3. Therefore,
it is clear that the symmetry of CF given by (1) is not the same
 as in the experimental data. The effective charge in the CF is
 1/3 but it is 1/2 in the data. There is no evidence of 
attachment of two flux quanta to one electron in the data.
Therefore, the theory of CF is not applicable to the 
experimental data of Kukushkin et al and they have made incorrect
 identification of data with CF. In short, all the efforts made
 to identify the CF in the experimental data have failed.
The relation which determines the effective field by attaching 
two flux quanta to one electron has to be abolished. This means
 that fluxes are not attached to the electrons. Therefore, even 
number of fluxes, such as two, are not attached to the electrons.
 When we examine the experimental data, there is no feature which 
can be identified with ``even number" of flux quanta attachment.

\noindent{\bf 3. Correct Theory}.\\

We show below that the use of our theory [9-13] solves the problem 
correctly. The quantum Hall effect is explained by suggesting that
 in a large field, the electron acquires large angular momentum 
and the effective charge is determined from the modifications of 
the Bohr magneton. The charge is coupled to the spin and there 
are two series which determine the effective charge.
 One series is obtained with positive sign of the spin and the other with
 negative
 sign as in $j= {\it l} \pm s$. One of the series is
 ${\it l/(2l+1)}$
and the other is ${\it l+1/(2l+1)}$. When we substitute, 
{\it l}=2, the two series give 2/5 and 3/5 and these are 
the particle-hole Kramers conjugate states. For ${\it l}$=3,
 the fractions are 3/7 and 4/7 and these are also the Kramers
 conjugate states. The values of 3/5 and 3/7 as predicted by us
 are clearly seen in the experimental data of ref.7. All of the 
values which we tabulated [9] theoretically in 1986 are the 
same
 as those given by the experimentally measured values of 
Fig.18 of St\"ormer[14]. Therefore, we regard our theory as correct.
 We have made considerable study of the data and find that all
 of the experimental data found in the Phys. Rev. Letters 
agrees with our calculations.\\

    In recent years, large values of the effective fractional
 charges have been obtained by spectroscopic methods. In one case, 
the sample is illuminated with a red light and then this light is
switch off and subsequently the resistivity is recorded[15]. In
 this case,  the resolution is considerably improved so that the 
fractions like 11/2 and 16/2 become visible. Such large values 
are not due to attachment of flux quanta to an electron. There is
no mechanism of gluing the flux quanta with electrons and there
 is no experimental evidence of flux quanta coming off the 
electrons.  This is a bit too far fetched idea and hence can be
 safely discarded. Willet et al [16] also find that as higher
 Landau levels are traversed, the validity of the CF picture is 
questionable.
However, they report that there is a filled Fermi sea of CFs at 5/2.
It turns out that at 5/2, there may be a Fermi sea but it does 
not prove that flux quanta are attached to electrons. In any case,
 there is no evidence of even number of flux quanta attaching
 to electrons. Auslaender et al [17] have performed the experiments
 with parallel wires and found that flux is quantized in units of $hc/e$
 but there is no evidence of attachment of flux quanta to electrons.
Therefore, the correct interpretation involves flux quantization and 
not flux attachment.\\
\\
\noindent{\bf 4. Conclusions}.\\

We have shown clearly that fluxex are not attached to the electrons
 and there is no need of composite fermions what so ever. The claim
 made by Kukushkin et al to have seen the CFs is completely incorrect 
and unjustified. Therefore, we observe that (a) the expression (1)
 for B$^*$ is incorrect because it does not obey the time reversal
 invariance and it does not quantize the flux correctly. (b) The 
fractional charge which the attachment of two fluxex to one electron 
produces is 1/3 in the CF whereas it is 1/2 in the experiment[7]. (c)
 There is no experimental evidence to support the suggestion of attachment
 of ``even number" of flux quanta to an electron and 
the claim made by the experimentalists is incorrect.
\newpage
\noindent{\bf References.}
\begin{enumerate}
\item J. K. Jain, Phys. Rev. Lett. {\bf63}, 199 (1989).
\item K. Park and J. K. Jain, Phys. Rev. Lett. {\bf81}, 4200 (1998).
\item R. K. Kamila, X. G. Wu and J. K. Jain, Phys. Rev. Lett. {\bf76}, 1332 (1996).
\item K. Park and J. K. Jain, Phys. Rev. Lett.{\bf80}, 4237 (1998).
\item J. K. Jain and R. K. Kamila, Int. J. Mod. Phys. {\bf11},2621 (1997).
\item K. Park and J. K. Jain, Phys. Rev. Lett. {\bf84}, 5576 (2000).
\item I. V. Kukushkin, J. H. Smet, K. von Klitzing and 
W. Wegscheider, Nature {\bf413}, 409 (2002).
\item K. N. Shrivastava, cond-mat/0201232.
\item K. N. Shrivastava, Phys. Lett. A{\bf113}, 435 (1986);115, 459(E)(1986).
\item K. N. Shrivastava, Mod. Phys. Lett. B {\bf13},1087 (1999).
\item K. N. Shrivastava, Mod. Phys. Lett. B{\bf14}, 1009 (2000).
\item K. N. Shrivastava, Superconductivity: Elementary Topics, 
World Scientific, New Jersey 2000.
\item K. N. Shrivastava, in Frontiers of Fundamental Physics 4, 
edited by B. G. Sidharth and M. V.
 Altaisky, Kluwer Academic/Plenum Pub. New York 2001.
\item H.L. St\"ormer, Rev. Mod. Phys. {\bf71}, 875 (1999).
\item K. B. Cooper, M. P. Lilly, J. P. Eisenstein, T. Jungwirth, 
L. N. Pfeiffer and K. W. West, Solid State Commun. {\bf119}, 89 (2001).
\item R. L. Willett, K. W. West and L. N. Pfeiffer,
 Phys. Rev. Lett. {\bf88}, 066801 (2002).
\item O. M. Auslaender, A. Yacoby, R. de Picciotto, K. W. Balwin, 
L. N. Pfeiffer and K. W. West, Science {\bf295}, 825 (2002).
\end{enumerate}

\end{document}